# Modified Confinement Model for Size Dependent Raman Shift and Linewidth of Silicon Nanocrystals

**Sanjeev K. Gupta and Prafulla K. Jha**[#]

Department of Physics, Bhavnagar University, Bhavnagar, 364 022, India

***Abstract***

A modified phonon confinement model considering the size distribution, an improved phonon dispersion curve and a confinement function is developed for the calculation of size dependent Raman spectra of the silicon (Si) nanocrystals. The model is capable in simultaneous calculation of the Raman shift, intensity and linewidth. The calculated size dependent redshift and linewidth of Raman spectra are in good agreement with the available experimental data in literature and better than previously reported theoretical results. The rapid rise in the redshift and linewidth for relatively smaller Si nanocrystals are well reproduced. The asymmetric behavior of Raman spectra is also obtained from the present model.

[#]Corresponding author: Prafulla K. Jha, prafullaj@yahoo.com; pkj@bhavuni.edu

## INTRODUCTION

In recent years the finite size effects on the Raman spectrum of a nanocrystal silicon have attracted considerable interest [1, 2] due to the role of Si in the current semiconductor industry and potential technological applications [3-5] resulting from the optical properties of Si quantum dots. However, most of the efforts are dedicated to interpret the experimental Raman spectra particularly in terms of confinement effect, which in 0D systems are difficult to evidence experimentally by Raman shift in contrast to 2D and 1D nanostructures. The quantization effect which is govern by many factors such as fluctuation in size, orientation and shape contributes in broadening of phonon peaks. Therefore, the majority of the published studies refer phonon confinement as the main mechanism responsible for the observed size-dependant Raman feature and a phonon confinement model (PCM) [6-11] has been proposed to interpret the Raman spectra. However, there are some reports on the limitations of the phonon confinement model (PCM) [3, 8-12]. To obtain a good interpretation of the experimental data many other models such as the microscopic force model [4], the bond polarizable model [13-14] and the spatial correlation model [15] have been reported. But these models do not satisfactorily describe the phonon confinement and are fail in explaining the experimental data particularly the shift and linewidth of the peak simultaneously. The widely used phonon confinement model is capable in describing the Raman peak shift and it's broadening in confined system by considering the breaking of $\mathbf{q}=0$ selection rule of infinite periodicity in the case of Raman scattering. For the spatially confined modes, the phonons away from $\mathbf{q}=0$ (i.e. $\mathbf{q}\neq 0$) also takes part in the scattering and results in the shift and broadening of the Raman peak [8-10, 12]. A suitable phonon confinement function is selected to correlate the size of the nanoparticles with Raman peak and its broadening alongwith the taking account of $\mathbf{q} \neq 0$ phonons with decreasing size. From experimental point of view, the Raman spectrum of Si nanocrystals exhibits a peak at 521 $cm^{-1}$ of Si bulk crystal, due to the optical phonon dispersion curves, with a linewidth $\Gamma$=3 $cm^{-1}$. However, the position of this peak and its broadening varies with size of the nanocrystals.

The models [4-8, 13-14] including the many version of phonon confinement model [9-12] failed in interpreting completely the size dependent Raman spectra perhaps due to the unproper selection of phonon dispersion curve and confinement function and neglect of size distribution function [8, 12]. It is expected that the inclusion of proper dispersion curve, confinement function and size distribution function may improve the understanding of phonon confinement and hence the interpretation of size dependent Raman spectrum. Recently, Faraci *et al* [12] have calculated the micro-Raman spectra in silicon nanocrystals by using a phonon confinement model. The calculated frequency shift and line broadening of Raman spectra compare well with the experiment for larger size but failed in the case of smaller size nanoparticles. The model is unable to reproduce the redshift as well as a rapid increase of redshift around 2-3 nm. The situation is similar for the linewidth. They emphasized for redefining the phonon dispersion curve and/or the confinement function alongwith the inclusion of size distribution function in the phonon confinement model for the better results. The expression for the dispersion curve used by them do not reproduce the actual dispersion of the phonon branch for which it has been used particularly at end points. The expression even fail to reproduce the exact experimental value of 521 $cm^{-1}$ at zone centre.

Motivated from the above fact, to study the size dependant Raman shift and linewidth of Si nanoparticles a modified phonon confinement model is proposed. The present model considers a new confinement function of Si nanocrystals, a suitable dispersion curves and the log-normal size distribution function. The selection of suitable dispersion curve is important due to high anisotropic nature of phonon dispersion curves of Si [7, 17], the size distribution has its importance for the smaller particle as has been pointed by Dièguez *et al* [18] in the case $SnO_2$ nanoparticles. The importance of phonon dispersion curves for the asymmetric broadening and shifting of Raman peak is already pointed out for the zinc and thorium oxide by Arora *et al* [16]. The expressions for dispersion curve used in the models are considered from its bulk dispersion curves. Most of the models so far used to understand the phonon confinement within the dot and hence for the qualitative description of the size dependent Raman shift considers arbitrary confinement function and expression for dispersion curves.

## PHONON CONFINEMENT EFFECT

Due to the Heisenberg uncertainty principle, the fundamental $\mathbf{q} \sim 0$ ($\mathbf{q}$ is the phonon wave vector) Raman selection rule is relaxed for a finite size domain, allowing the participation of phonon away from the Brillouin zone center, also contribute to the Raman lineshape. The phonon uncertainty goes roughly as $\Delta\mathbf{q} \sim 1/d$, where '*d*' is the quantum dot diameter. Consequently, the Raman spectrum [9-10, 12] is calculated by the following integral over the momentum vector $\mathbf{q}$,

$$I(\omega) \propto [n(\omega)+1] \int |C(\mathbf{q})|^2 L(\omega, \mathbf{q}) d\mathbf{q} \quad (1)$$

Where the integral is extended to the entire first Brillouin zone, $\omega$ is the Raman frequency, $n$+1 is the Bose-Einstein factor, *C(**q**)* are Fourier coefficients of the phonon wave function and *L(ω, **q**)* is the Lorentzian function related to the phonon dispersion curve. The Raman scattering intensity, $I(\omega)$, for nanocrystals with the size distribution ρ(L) is given by [8, 19],

$$I(\omega) \propto \int \rho(L) dL \int \left( \frac{|c(\mathbf{q})|^2}{(\omega - \omega(\mathbf{q}))^2 + \left(\Gamma_0 / 2\right)^2} \right) d^3\mathbf{q} \tag{2}$$

Where $\Gamma_0$ is the FWHM of the Raman line of bulk, $\omega(\boldsymbol{q})$ is the phonon dispersion curve, *c(**q**)* is the Fourier coefficients of the vibrational weighting function expanded in the Fourier integral and $\rho(L)$ is the log-normal size distribution, which corresponds to the density of vibrational states, since each particle vibrates with a frequency that is inversely proportional to its diameter 'D'. Initial wave vector $\mathbf{k_n}$ can contribute to the photon scattering process with an effective wave vector **q** with the spread allowed by the Heisenberg uncertainty principle around the initial value of $\mathbf{k_n}$. It can be written as $\mathbf{k_n} = n\pi/D$ with D as the size of spherical nanocrystal, $n$=2, 4, 6…$n_{max}$ ($n_{max}$ is equal to the maximum integer smaller than 2D/$a$, and $a$=0.543 nm is the Si lattice parameter) and $\rho(L)$ representing the log-normal distribution for the size of particles can be defined as,

$$\rho(L) = \exp\left( \frac{-\log\left( D / D_0 \right)^2}{2\sigma^2} \right) \tag{3}$$

And the parameters 'σ' measures the distribution width. $D_0$ is the diameter that corresponds to the maximum of the distribution. The Fourier coefficients of the vibrational weighting function *c(q)* for spherical particles in the present case is considered as [9-10],

$$|c(\mathbf{q})|^2 \approx \left(\frac{1}{\left(16\pi^4 - d^2\mathbf{q}^2\right)}\right)^4 \tag{4}$$

In the calculation of Raman intensity of Si nanocrystals, we have used $\Gamma_0$= 3 $cm^{-1}$ and ω= 521 $cm^{-1}$. In most of the earlier studies the dispersion curves for the calculation of Raman intensity for Si nanocrystals and several others [6-8, 12, 19] are considered one which is easy to be presented or fit to the smoother curves. In the present calculation we have calculated the Raman intensity by using the dispersion curves of the phonon branch in **Γ-X** direction of Brillouin zone represented by the analytical form [8, 17-18, 20],

$$\omega(\mathbf{q}) = \sqrt{A + BCos(\mathbf{q}a/2.6)} \tag{5}$$

where, A is $1.714\times10^5$ $cm^{-2}$ and B is $1.2\times10^5$ $cm^{-2}$, with **q** in the range of ($n\pi+1/D$) to ($n\pi-1/D$). The dispersion curve represented by the Eq (5) is plotted in Fig. (1) alongwith the data of ref. [12, 17]. The Fig. 1 reveals that the present function produces the dispersion curves better than the previous used function [12].

## RESULTS AND DISCUSSION

In Fig. 2, we present the calculated Raman spectra by using Eq. (2) for the different sizes of Si nanocrystals. The figure depicts that at smaller size, there is a significant shift and broadening of the spectra, while for the particles of larger sizes the peak is not shifted significantly and approaches the bulk value of 521 $cm^{-1}$ alongwith its natural width of 3 $cm^{-1}$. The asymmetric lineshape (tailing towards lower Raman shifts) in the Raman spectra is also well reproduced for the nanocrystalline Si similar to the previous studies [6-8, 12]. In Fig. 3, we plot the size dependent Raman frequency redshift for Si nanocrystals. This figure also includes the theoretical and experimental data reported from literature for the comparison. The Fig. 3 reveals the close agreement with the experiments in case of both smaller and larger particles. As can be seen from Fig. 3, except the present model none of the earlier models is able to satisfactorily produce the simultaneous observed a higher shift and a rapid increase around 2-3 nm size Si nanoparticles. This is due to the fact that the present model includes the effect of size distribution and perhaps a most appropriate dispersion curves alongwith the confinement function.

The present model is also use to calculate the linewidth of the Raman spectrum. In Fig. 4, we presents the size dependent line broadening for Si nanocrystals alongwith the experimental and theoretical data from literature for comparison. An instrumental linewidth FWHM of 3.0 $cm^{-1}$ is assumed. The present model is in general good agreement with the experimental data and better than the previous calculation [12] similar to the redshift. The rapid rise around 2 nm is also produced well. The slight variation in magnitude which is probably the singular failure of the approach cannot be a drawback particularly in the case of large scatter experimental data.

## CONCLUSION

In summary, we have presented a modified phonon confinement model which includes a new confinement function, size distribution and a dispersion curves representing accurately the most dispersive behavior of phonon branch to explain the size dependent Raman spectra for Si nanocrystals. The results obtained using the present model is in general fair agreement with the experimental data. The success of present calculation supports the idea of inclusion of size distribution in the phonon confinement model. The present calculation is also able to reproduce the asymmetric behavior of Raman spectra. Finally, in our opinion the results of the present work would be useful in the way to get a better understanding of phonon confinement in nanoparticles.

**ACKNOWLEDGEMENTS**

The authors gratefully acknowledge Dr. A K Arora for fruitful discussion. The financial assistance from DAE-BRNS and Department of Science and Technology, Govt. of India are highly appreciated.

## Figure Caption

**Figure 1:** Plot of dispersion curves. Symbols indicate the experimental data [17].

**Figure 2:** Calculated Raman spectrum with Si nanocrystal size; spectra are normalized to the same height, taken equal to unity. The natural linewidth full width at half maximum (FWHM) $\Gamma = 3\ cm^{-1}$.

**Figure 3:** Raman line shape of Si as a function of particle size. Various symbols indicate the experimentally and calculated data for comparison.

**Figure 4:** Plot of the Raman linewidth as a function of the Si nanocrystal size. The natural linewidth full width at half maximum (FWHM) $\Gamma = 3\ cm^{-1}$. Experimental and other calculated data from the literature are displayed for comparison.

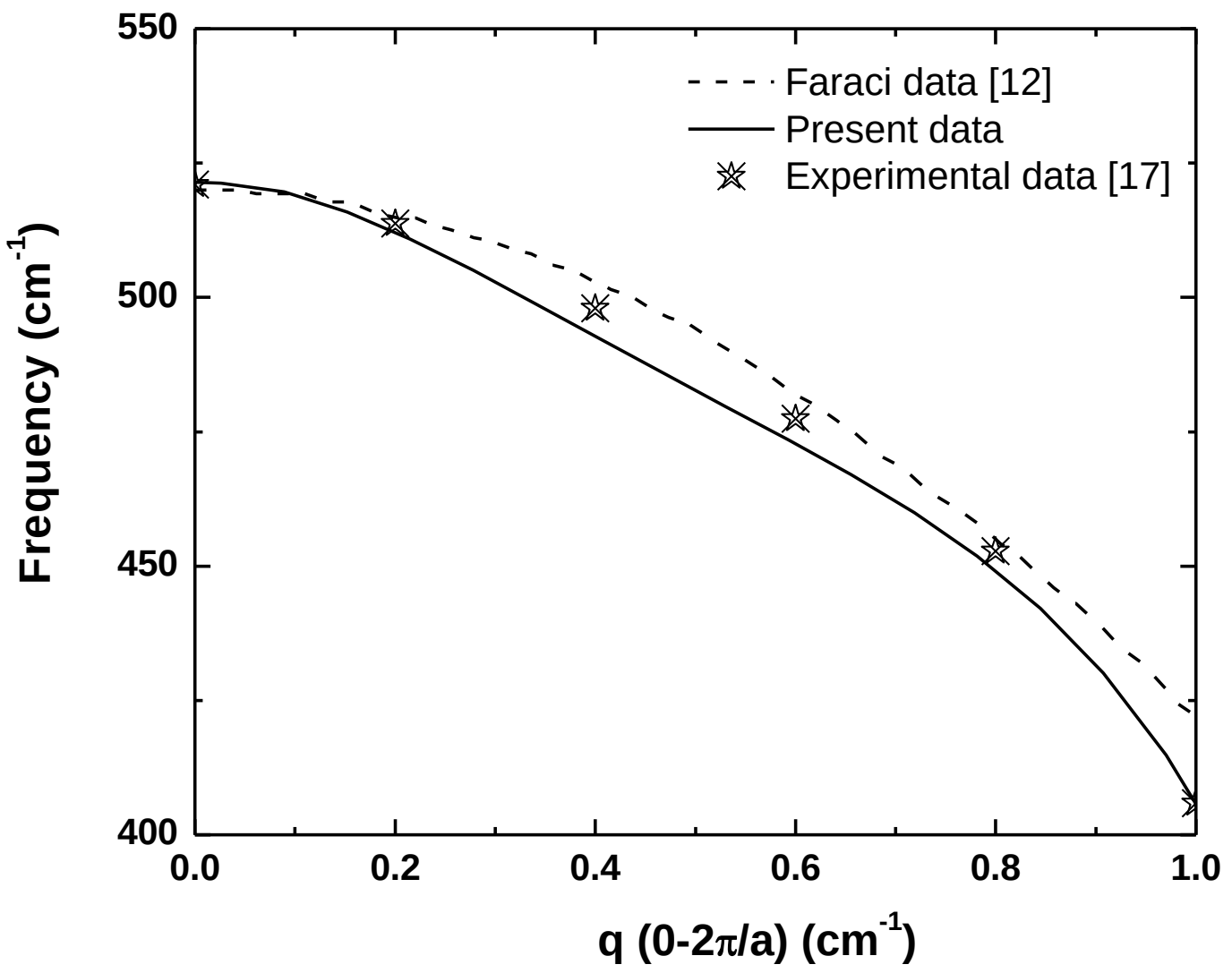

**Fig. 1** Gupta *et al*.

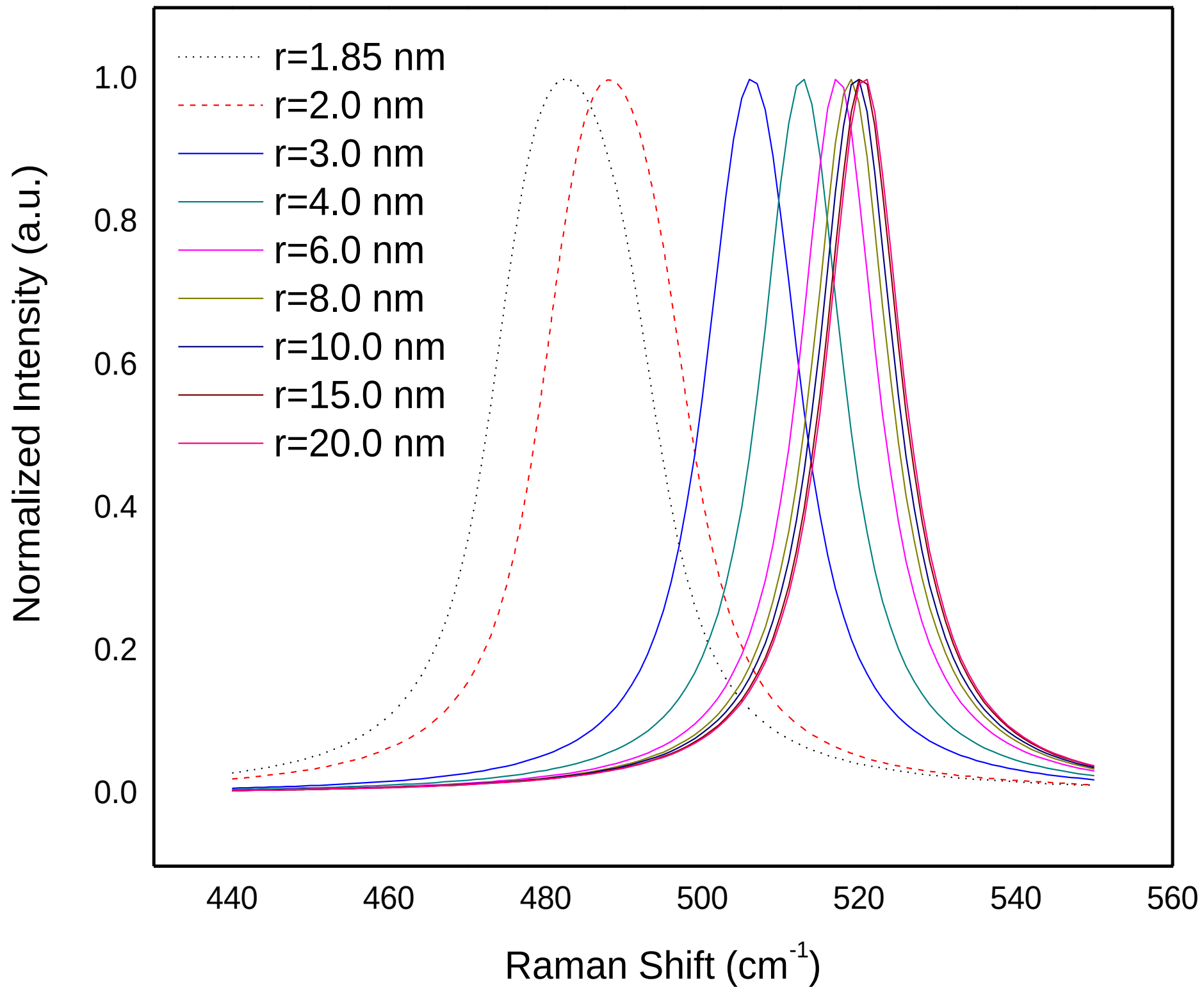

**Fig. 2** Gupta *et al*.

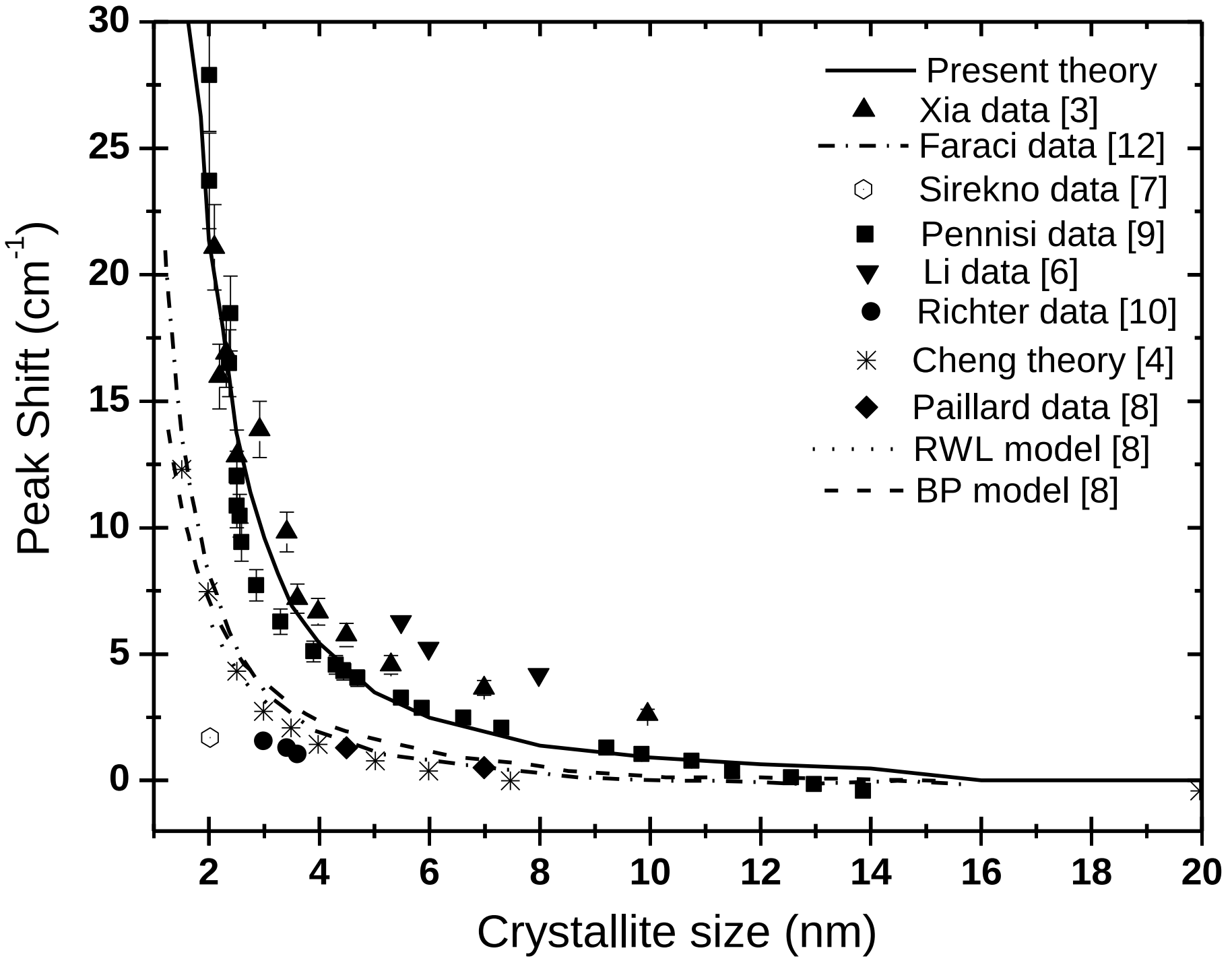

**Fig. 3** Gupta *et al*.

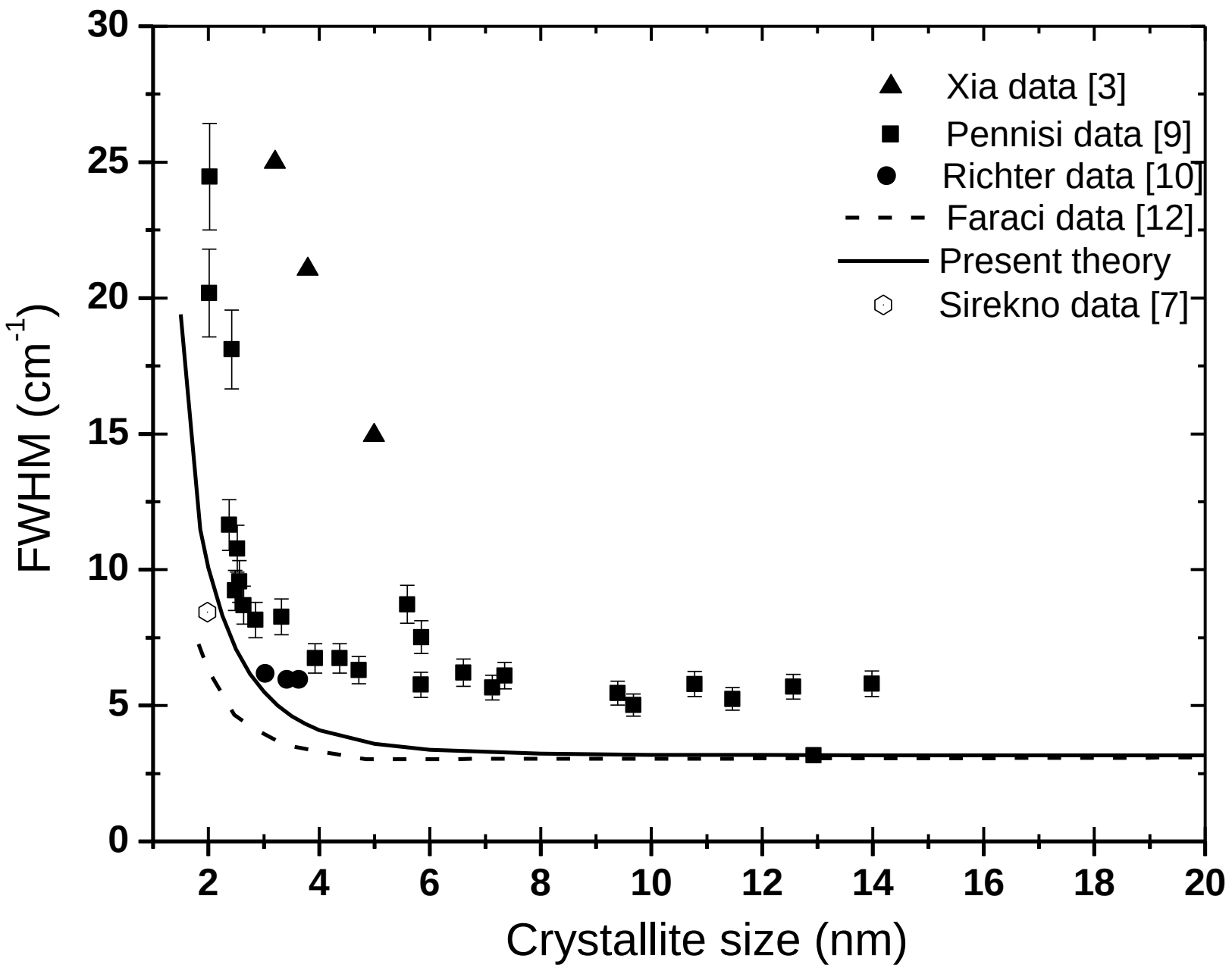

**Fig. 4** Gupta *et al*.